\documentclass[aps,prl,reprint,groupedaddress]{revtex4-1}

\usepackage{graphicx}
\usepackage{graphics}

\begin{document}

\preprint{}

\title{Correlations beyond the horizon}


\author{Daniel Golubchik}
\email[danielg@tx.technion.ac.il]{}
\author{Emil Polturak}
\author{Gad Koren}
\affiliation{Department of Physics, Technion - Israel Institute of Technology,
Haifa 32000, Israel}


\date{\today}

\begin{abstract}
We have imaged spontaneously created arrays of vortices (magnetic
flux quanta), generated in a superconducting film quenched through
its transition temperature at rates around $10^9 K/s$. Spontaneous
appearance of vortices is predicted by Kibble-Zurek and by
Hindmarsh-Rajantie models of  phase transitions under
non-equilibrium conditions. Differentiating between these models
requires a measurement of the internal correlations within the
emerging vortex array. In addition to short range correlations
predicted by Kibble and Zurek, we found unexpected long
range correlations which are not described by any of the existing
models.
\end{abstract}


\maketitle

Any physical system undergoing a phase transition within a finite
time interval is necessarily driven out of equilibrium. One model
that describes the dynamics of such phase transitions is the
Kibble-Zurek scenario. The model was first suggested by Kibble
\cite{Kibble:1976,Davis:2005} for cosmological phase transitions
which occurred in the early universe. Kibble proposed that fast
cooldown of the universe, through a critical temperature $T_c$,
leads to formation of initially isolated domains of a new ordered
phase. The typical size of  a domain, $\hat{\xi}$ , inside which
the emerging order is coherent, is a product of the speed of light and time needed to complete the phase transition. Consequently, different domains are uncorrelated due to causality. As these disjoint
domains coalesce, the mismatch of the order parameter between
different regions leads to the appearance of topological defects. The domain size determines both the typical distance and the correlation length between topological defects.
Zurek \cite{Zurek:1985} proposed terrestrial tests of this model
by examining analogous situations in condensed matter systems
having the same symmetry of the order parameter. The analogous
"speed of light" in condensed matter systems is the velocity of
propagation of the order-parameter.
 Since, some aspects of the
Kibble-Zurek (KZ) model has been tested in a variety of physical
systems, including liquid helium \cite{Bauerle:1996,Ruutu:1996},
liquid crystals \cite{Ray:2004}, superconductors
\cite{Maniv:2003,Kirtley:2003} and Josephson junctions
\cite{Monaco:2002,Carmi:2000}. Thus, the Kibble-Zurek model is a
universal theory of defect formation whose applications range from
phase transitions in grand unified theories of high-energy physics
to phase transitions observed in different condensed-matter
systems. An alternative mechanism of spontaneous vortex formation in
superconductors was proposed by Hindmarsh and Rajantie
\cite{Hindmarsh:2000,Rajantie:2009} (HR). According to this
mechanism, thermal fluctuations of the magnetic field freeze
inside the superconductor during the transition, creating domains
of magnetic flux with the same polarity and characteristic size.
In contrast, the KZ model predicts only short range correlations,
with neighboring defects having different topological charge. In a
superconductor, where the topological defects are vortices
carrying a quantum of magnetic flux, this means that adjacent
vortices should have a different polarity. The presence of
spontaneously generated topological defects was observed in
several experiments. Regarding the more sensitive testing of
correlations predicted by these models, there is only one
experiment on liquid crystals in which an array of topological
defects was actually imaged. However, the amount of data was
insufficient to detect correlations beyond nearest neighbours. The
objective of our experiment is to decide between these two models
by imaging spontaneously formed arrays of vortices in a
superconductor and measuring their correlations.

Investigating the spatial distribution of the vortex array
requires a technique capable of imaging a relatively large area
with a micrometer resolution. Furthermore, the statistical nature
of the problem requires accumulating many such images. The only
practical method to achieve both these goals is Magneto-Optical
(MO) imaging. Resolving single vortices however requires pushing
the technique to its limits, and until now only one group has
achieved this goal \cite{Goa:2003}. We have developed a high
resolution Magneto-Optical system specifically for this
experiment. Our system is described in detail
elsewhere\cite{Golubchik:2009}. Briefly, having the
magneto-optical indicator as close as possible to the
superconductor's surface and a cryogenic design suppressing
vibrations allowed us to achieve the best spatial resolution
($0.8\mu m$) demonstrated so far by this technique
\cite{Golubchik:2009}. The superconductor sample consists of a
$200 nm$ thick Niobium film deposited on a sapphire substrate. The
film was prepared by DC magnetron sputtering and it is
superconducting below $8.9K$. The film is patterned into small
squares of $200\mu m$ across, to ensure homogeneous illumination
by the heating laser pulse and avoid thermoelectric currents
inside the sample. On top of the Nb film, we deposited a $40 nm$
layer of $EuSe$ which serves as the Magneto-Optic sensor. Due to
the huge magneto optical Kerr effect in EuSe the polarization
plane of linearly polarized light is rotated if local magnetic
field is present. By mapping this rotation at each point of the
image, we get an image of the magnetic field directly above the
surface of the superconductor. The typical lateral size of a
vortex in a Nb film is about $100 nm$ \cite{Gauzzi:2000}, so the
images of individual vortices are diffraction limited. In our
setup, observation of individual vortices is possible over a large
field of view ($100\times100 \mu m^2$) using relatively short
integration time ($10 s$), allowing us to acquire hundreds of
images during an experimental run.

From our previous experiments \cite{Maniv:2003}, we know that
extremely high cooling rates are essential for spontaneous
generation of a measurable amount of vortices. No less important,
fast cooling to low temperatures (far below $T_c$) traps the
vortices on pinning centers, preventing mutual annihilation of a
vortex and an anti-vortex. High cooling rates are achieved in the
following way: first, the superconducting film is heated above
$T_c$ by a short laser pulse. The $200 nm$ thick film is deposited
on a sapphire substrate, which is transparent at the wavelength of
the laser. Hence, only the film heats up, while the $1 mm$ thick
substrate remains near the base temperature. At the end of the
heating pulse, the heat from the film escapes via ballistic
phonons into the cold substrate, which has a thermal mass of
$\sim1000$ larger than that of the film and acts as a heat sink
during the cooldown. At low temperatures involved (below $14K$),
phonon scattering is small enough so that the heat transfer from
the film into the substrate is ballistic. The timescale of the
heat transfer is much shorter than the length of the laser pulse.
Therefore, the cooling rate depends only on the decay time of the
laser pulse. We used pulse shaping techniques to change the decay
time. In this way two cooling rates of $4\cdot10^8$ and
$2\cdot10^9 K/s$ were achieved. The cooling rates were measured
using $GeAu$ thin film resistive bolometer. The conductivity of
the $GeAu$ film, which is temperature dependent, was measured vs.
time during the laser pulse. Finally, at these cooling rates, the
mean inter-vortex spacing predicted by the models is larger than
our optical resolution, so that we should be able to observe the
entire vortex array.

\begin{figure}
\includegraphics [width=3in]{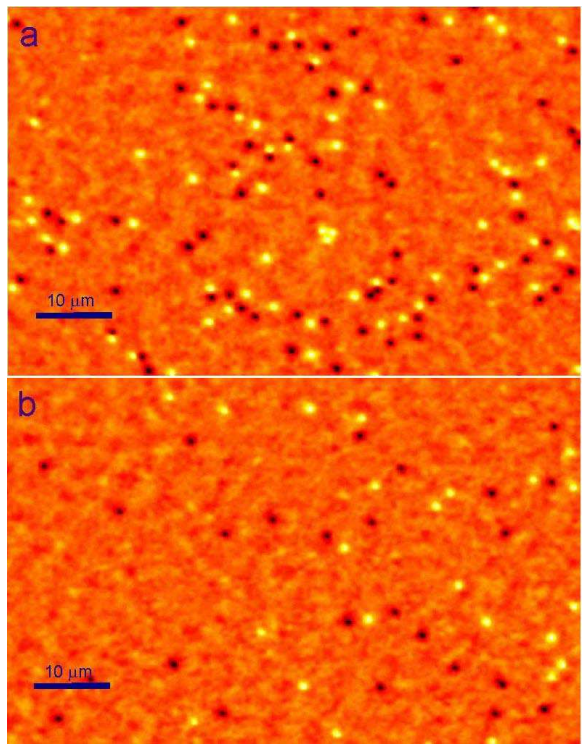}
\caption{Spontaneously created vortices. Typical images of
spontaneously created flux lines in a superconductor cooled at: a)
$2\cdot10^9 K/s$, and b) $4\cdot10^8 K/s$. The intensity is
proportional to the local magnetic field. Bright and dark spots
represent vortices with opposite polarity. \label{spont}}
\end{figure}

Typical images of spontaneously generated vortex arrays are shown
in Fig.1. During these measurements, the system was carefully
shielded from external magnetic fields, and the measured asymmetry
between the density of positive and negative vortices is less than
$1\%$. It is clear that the vortex arrays appear random. On
average, the two cooling rates we could use, $4\cdot10^8$ and
$2\cdot10^9 K/s$, produced a density of vortices of $6\cdot10^5
cm^{-2}$ and $1.3\cdot10^6 cm^{-2}$ respectively. Those densities
are almost $2$ orders of magnitude lower than expected from the KZ
model. This discrepancy was already noticed in our earlier
experiments\cite{Maniv:2003}. However, the scaling of the density
with the cooling rate is consistent with the KZ model
(proportional to the square root of the cooling rate). One
potential reason for the low density could be the mutual
annihilation of nearby vortices having opposite polarities. The
closer vortices are to each other, the stronger the attractive
force between them. Vortices are prevented from mutually
annihilating by pinning forces, which however do not depend on the
distance. Therefore a critical distance between vortices with
opposite polarity exists, below which pairs of vortices will
overcome the pinning force, merge and annihilate. We determined
this critical distance by repeating the experiment under various
external magnetic fields (Fig. 2). When the density of the
vortices due to external field is such that the distance between
them is less than critical, all the vortices having an opposite
polarity should annihilate. This critical length turns to be $1
\mu m$ or less, significantly smaller than the scale of short
range correlations. Hence, annihilation should not affect any of
the correlations observed in our experiment.

\begin{figure}
\includegraphics[width=3in]{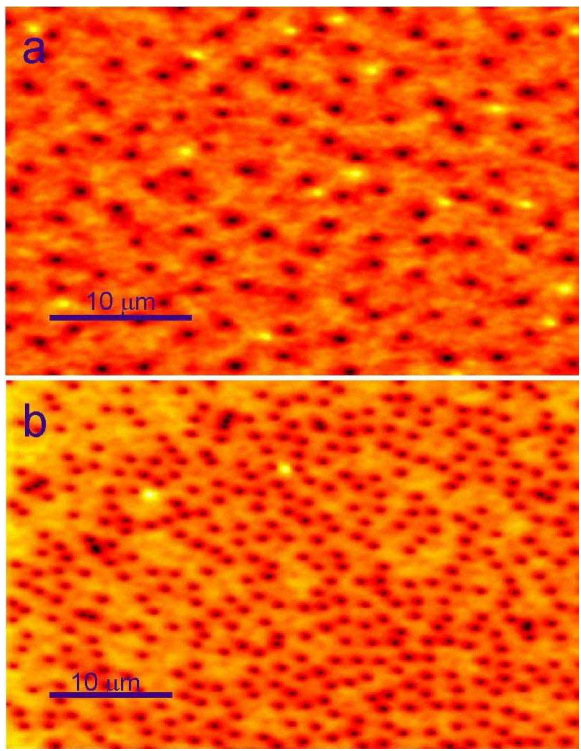}
\caption{Images of magnetic flux above the superconductor cooled
in the presence of external magnetic field. The external field in
(a) is $0.2 mT$ and in (b) is $4$ times larger, $0.8 mT$. The
cooling rate is $10^9 K/s$ in both panels. The majority of the
vortices (dark spots) are associated with the external magnetic
field. Bright spots are vortices with polarity opposite to the
external magnetic field. Due to annihilation, at higher external
fields the number of vortices with opposite sign is reduced.
\label{anihil}}
\end{figure}

We used images like Fig.1 to determine the correlations between
the vortices. In order to increase the statistical ensemble, the
correlation function was averaged over $170$ such images, with
$12,000$ vortices in total. We first show the density-density
correlation function, irrespective of the vortex polarity. This
function is defined as $D(r-r^\prime)=<\rho(r)\rho(r^\prime)>$
where $\rho(r)$ is the local vortex density. The value of
$\rho(r)$ is taken to be $1$ at the location of a vortex
regardless of its polarity and $0$ elsewhere. According to the KZ
model, if a vortex is created at some vertex between different
domains, the probability to find another vortex at the nearest
neighbour vertex is higher by $33\%$ than the average.
Consequently, $D(r)$ should show a peak at the characteristic
nearest neighbour distance. The correlation function calculated
from our data is presented in Fig.3. There is indeed a peak at
$r\approx3 \mu m$, a strong evidence for the short range
correlations predicted by the KZ model. Beyond the peak, the
correlation function decays towards a constant value. The nearest
neighbour distance is also the best estimation for a typical
domain size $\hat{\xi}$.

\begin{figure}
\includegraphics[width=3in]{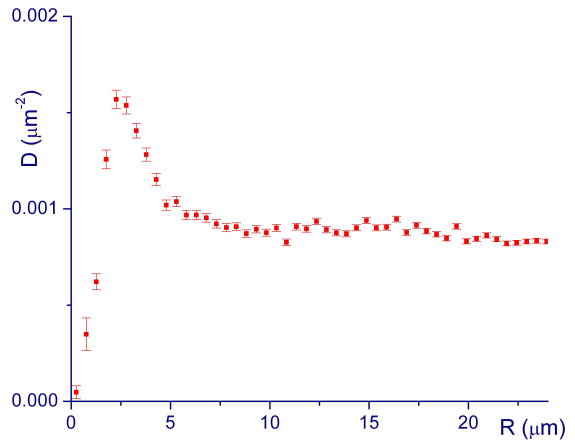}
\caption{Density-density correlation function $D(r)$. The error
bars reflect the statistics. The peak at $3 \mu m$ is evidence for
strong short range correlations predicted by the KZ model.
\label{dens}}
\end{figure}

Next we calculate the correlation function taking into account the
polarity of each vortex. The vortex-vortex correlation function is
defined as $G(r-r^\prime)=<n(r)n(r^\prime)>$ , with  $n(r)=1$ at
the location of a positive vortex, $-1$ at the location of
negative vortex and $0$ elsewhere. The KZ model predicts that
nearest neighbor vortices should have opposite polarities. This
should manifest itself as a negative peak in $G(r)$, which
according to theory \cite{Liu:1992} should decay exponentially to
zero ($G(r)\propto r^2exp(-r^2/\hat{\xi}^2)$. In the KZ model,
$\hat{\xi}$ is the only length scale, and both the decay length of
the correlations and the characteristic distance between vortices
should be $\cong\hat{\xi}$. In contrast, in the HR model
neighboring vortices should have the same polarity and so the peak
in the correlation function should be positive.

Fig.4 shows the vortex-vortex correlation function. The
correlation function was multiplied by r to emphasize long range
behavior. The nearest neighbour peak is negative, which indicates
that the KZ scenario is the dominant mechanism of vortex
formation. The solid line is a fit to the theory of
\cite{Liu:1992} with $\hat{\xi}=3$. It indeed appears that the
nearest neighbour distance (the peak in Fig.3) and the decay
length of the short range correlation (Fig.4) are the same.
However, at distances beyond the peak the theoretical correlation
function decays to zero, while the experimental correlations do
not. Surprisingly, the data show an oscillatory behavior which
decays as $\propto r^{-\alpha}$, with $\alpha=1\pm 0.5$. This
oscillatory behavior is well outside the error margins. Such
oscillations, if any, are within the noise in Fig.3. Hence the
oscillations in Fig.4 represent long range correlations in the
polarity of vortices rather than in their density. We found that
the wavelength of these oscillations decreases weakly with
increasing cooling rate as well as with applied magnetic field.
One possibility is that these long range correlations result from
some local inhomogeneities in the sample. To check this
possibility, we repeated the experiments at several different
locations on the film. We found that the correlations were
independent of the location. Further, local variations in the
local properties such as the value of $T_c$ could potentially
modulate the density of vortices which however is constant (fig.
3), but not the polarity of vortices which is what we see.

\begin{figure}
\includegraphics[width=3in]{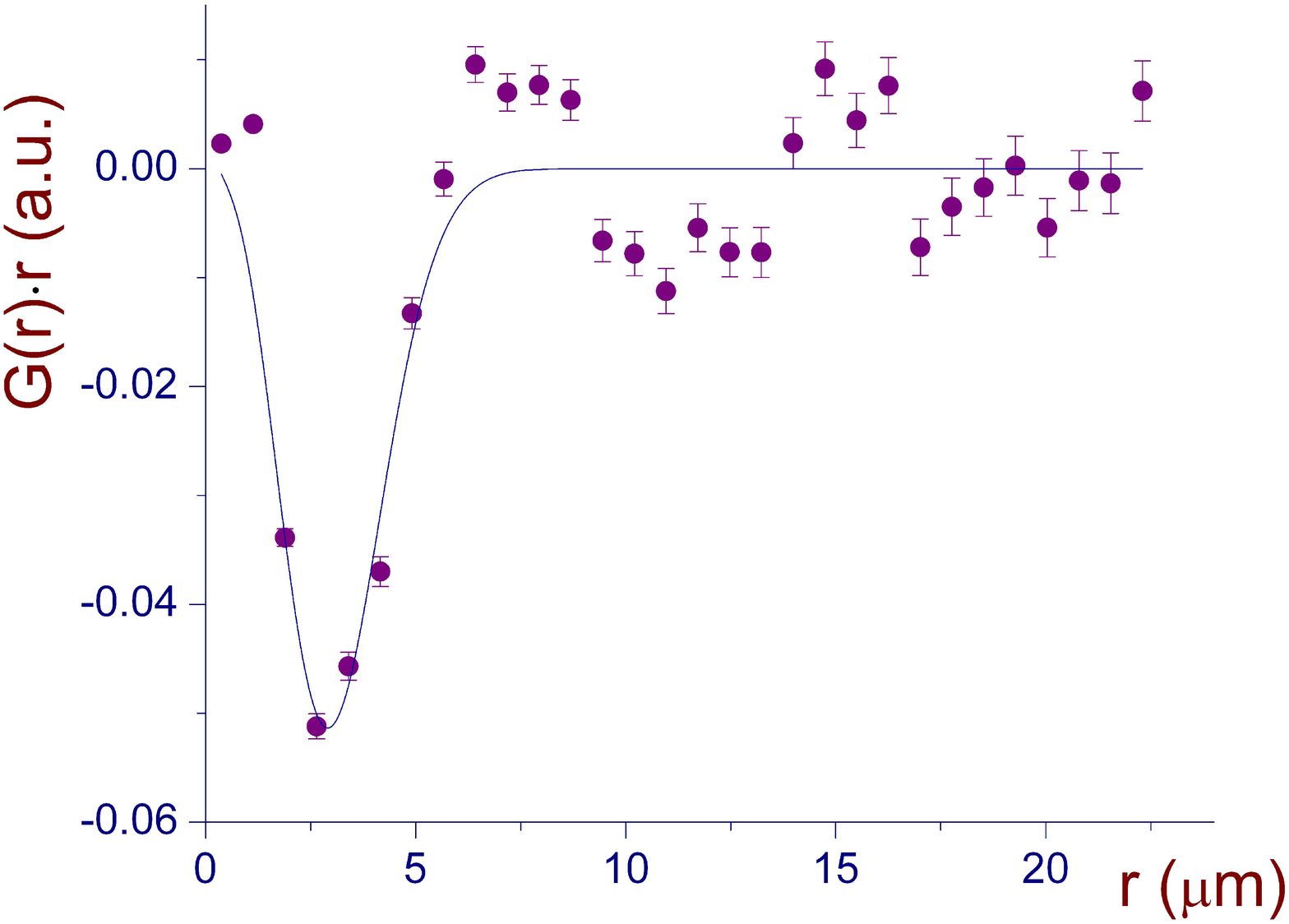}
\caption{First moment of the vortex-vortex correlation function
$G(r)$. The negative peak at short distance reflects
vortex-antivortex correlations predicted by KZ model. The blue
line is a fit to theory of \cite{Liu:1992}. \label{vortcor}}
\end{figure}

Long range correlations were predicted in the HR model
\cite{Hindmarsh:2000,Rajantie:2009}. However, these correlations
should decay as $r^{-4}$, instead of $r^{-1}$ which we observe and
the predicted domain size is on a $mm$ scale, $2$ orders of
magnitude larger than the few $\mu m$ which we see. In cosmology,
the size of a domain is determined by the speed of light. The
analogous speed in superconductors is the propagation velocity of
the order parameter \cite{Zurek:1985}. Close to $T_c$, this speed
is about $10^3 m/sec$. This speed determines the nearest neighbour
distance, the only correlation length in the KZ model. However,
vortices are also coupled to electromagnetic field, which
propagates with speed $c$. Such coupling may perhaps be linked
with the longer range correlations, which are currently a puzzle.

\begin{acknowledgments}
We thank S. Lipson and E. Buks for their contribution to this
experiment. We thank S. Hoida, L. Iomin and O. Shtempluk for
technical assistance. This work was supported in part by the
Israel Science Foundation and by the Minerva and DIP projects.
\end{acknowledgments}



\end{document}